\documentclass[12pt]{article}

\usepackage[utf8]{inputenc}
\usepackage{graphicx}
\usepackage{float}
\usepackage{bbm}
\usepackage{amsmath}
\usepackage{braket}
\usepackage{amsfonts}
\usepackage{amssymb}
\usepackage{appendix}
\begin{document}

\begin{center}

\textbf{\Large Probing Source and Detector NSI parameters at the DUNE Near Detector } 

\vspace{30pt}
 A. Giarnetti$^a$ and D. Meloni$^a$
\vspace{16pt}

\textit{$^a$Dipartimento di Matematica e Fisica, 
Universit\`a di Roma Tre\\Via della Vasca Navale 84, 00146 Rome, Italy}\\
\vspace{16pt}

\end{center} 

\abstract
We investigate the capability of the DUNE Near Detector (ND) to constrain Non Standard Interaction parameters (NSI) describing the production of neutrinos ($\varepsilon_{\alpha\beta}^s$) and their detection ($\varepsilon_{\alpha\beta}^d$). We show that the DUNE  ND is able to reject a large portion of the parameter space allowed by DUNE Far Detector analyses 
and to set the most stringent bounds from accelerator neutrino experiments on $|\varepsilon_{\mu e}^{s,d}|$  for wide intervals of the related phases. We also provide simple analytic understanding of our results as well as a numerical study of their dependence on the systematic errors, showing that the DUNE ND offers a clean environment where to study source and detector NSI. 
\section{Introduction}
Thanks to the increasing evidence of non-vanishing CP violation in the lepton sector \cite{Abe:2019vii}, the standard three-neutrino oscillation framework seems to be rather established; however, the precision on the mixing parameters is above the percentage level \cite{Esteban:2018azc,deSalas:2017kay} and this leaves room for effects not described by the standard physics. To catch the relevant impacts of possible new physics signatures in a model independent way, a useful approach relies on the employment of effective four fermion operators, the so called Non-Standard Interaction operators (NSI), that arise from the presence of heavy mediators \cite{Grossman:1995wx}-\cite{Bergmann:1999rz}.
If not diagonal in the flavor basis, they can affect the interactions between neutrinos and charged leptons and, in particular, influence neutrino oscillations; thus, we can distinguish among  three different scenarios:
\begin{itemize}
    \item the decaying particles that produce a neutrino of flavor {\it $\alpha$} associated to a charged lepton is also able to produce other neutrino flavors {\it $\gamma$}. Thus, at the source {\it s}:
    \begin{equation}
        \ket{\nu_\alpha^s}=\ket{\nu_\alpha}+\sum_{flavors}\varepsilon_{\alpha \gamma}^s\ket{\nu_\gamma}\,,
        \label{eqs}
    \end{equation}
    where $\varepsilon_{\alpha \gamma}^s$ is a $3 \times 3$ matrix of unknown coefficients describing the amplitude of the contamination of flavors other than $\alpha$;
    \item during their propagation, neutrinos oscillate and can interact with matter, developing an effective potential that modifies the vacuum oscillation probabilities. NSI effects add new contributions to the matter potential  parametrized in terms of coefficients $\varepsilon_{\alpha\beta}^m$;
    \item once in the detector {\it d}, a neutrino of flavor $\gamma$ can give rise to charged current interactions (CC) with nuclei or electrons which, in presence of NSI,  can produce a charged lepton  of a different flavor $\beta$. Thus:
    \begin{equation}
        \bra{\nu_\beta^d}=\bra{\nu_\beta}+\sum_{flavors}\varepsilon_{\gamma \beta}^d\bra{\nu_\gamma}\,,
        \label{eqd}
    \end{equation}
    where, as above, the coefficients  $\varepsilon_{\alpha \gamma}^d$ are describing the amplitude of the contamination of flavors other than $\beta$.
\end{itemize}
It is worth mentioning that the $\varepsilon_{\alpha\beta}^{d,s}$ previously introduced are effective couplings that receive contributions from four-fermion operators with different Lorentz structure:
\begin{eqnarray}\label{laggen}
  \mathcal{L}_{\rm NSI} &=& \mathcal{L}_{V \pm A} + \mathcal{L}_{S \pm P}
                            + \mathcal{L}_T,
\end{eqnarray}
where the operators interesting for this paper (that is the ones related to source and detector NSI's) are:
\begin{eqnarray}
  \mathcal{L}_{V \pm A} &\ni& 
  \frac{G_{F}}{\sqrt{2}} \sum_{f, f^\prime} \varepsilon^{f,f^\prime, V \pm A}_{\alpha\beta}
      \left[ \bar{\nu}_\beta \gamma^{\rho} (1 - \gamma^{5}) \ell_\alpha \right] \!
      \left[ \bar{f}^\prime \gamma_{\rho} (1 \pm \gamma^{5}) f \right] \nonumber\\
  \mathcal{L}_{S \pm P} &=&\nonumber
  \frac{G_{F}}{\sqrt{2}} \sum_{f, f^\prime} \varepsilon^{f,f^\prime, S \pm P}_{\alpha\beta}
      \left[ \bar{\nu}_\beta (1 + \gamma^{5}) \ell_\alpha \right] \!
      \left[ \bar{f}^\prime (1 \pm \gamma^{5}) f \right],
  \label{eq:NSI-Lagrangian-SpmP} \\
  \mathcal{L}_{T} &=&
  \frac{G_{F}}{\sqrt{2}} \sum_{f, f^\prime} \varepsilon^{f,f^\prime,T}_{\alpha\beta}
      \left[ \bar{\nu}_\beta \sigma^{\rho\tau} \ell_\alpha \right] \!
      \left[ \bar{f}^\prime \sigma_{\rho\tau} f \right].
  \label{eq:NSI-Lagrangian-T}
\end{eqnarray}
Here, $G_F$ is the Fermi constant, $\nu$ and $\ell$ are the neutrino and charged
lepton fields, and the $f$'s are the fermions participating  to the neutrino interactions. The strength of the NSI are encoded into the arbitrary complex matrices $\varepsilon$ which we will distinguish with the superscript $s$ or $d$ when the interaction takes place at the neutrino source or detector, respectively.
Since in the rest of the paper we will focus on the DUNE experiment, the indeces $f$ and $f^\prime$ are fixed to $f=u$ and $f^\prime=d$; thus, the source NSI's (for which $\ell_{\alpha}=e,\mu$) receive contributions from $V\pm A $ and $S\pm P$ operators (no P-odd part is present in the tensor operator, so $\mathcal{L}_{T}$ does not contribute) while 
the detector NSI's (for which again $\ell_{\alpha}=e,\mu$) receive the largest contributions only from the $V-A$ structure of the weak current \cite{Kopp:2007ne}.\\
With NSI taken into account, the parameter space describing neutrino oscillations is enlarged to incorporate, in the most general case, nine more complex parameters from the  $\varepsilon^s$ matrix, nine more complex parameters from
$\varepsilon^d$ and eight real parameters from the Hermitian matrix $\varepsilon^m$ \footnote{One parameter can be subtracted from the diagonal, bringing from nine to eight the number of independent matrix elements.}. In an accelerator experiment, $\varepsilon^s_{\tau\alpha}$ can be neglected since usually the $\tau$ production is absent or very small. For this reason,  in these experiments the number of involved source NSI parameters is reduced to six. 
It is clear that a simultaneous determination of all mixing parameters requires special care in what many correlations and  degeneracies  appear that  cannot  be completely broken  by  simplified analyses. However, it turns out that selected classes of neutrino experiments are sensitive to subsets of NSI parameters and can be used to constrain some of the entries of the $\varepsilon^{s,m,d}$ matrices. This is the case of solar neutrino experiments\footnote{Notice that the NSI parameters  that affect neutrino oscillations are combinations of those entering the Lagrangian describing the interaction processes. We assume here   that the quoted bounds  directly  apply to $\varepsilon^{s,m,d}$.}, where 90\% confidence level (CL) bounds on $|\varepsilon_{{e\mu},{e\tau},{\tau \tau}}^m|\sim {\cal O}(0.1)$ \cite{Coloma:2017ncl,Khan:2017oxw,Liao:2017uzy} and on $|\varepsilon_{{e e},{\tau \tau}}^d|\sim {\cal O}(10^{-1}-10^{-2})$ \cite{Bolanos:2008km,Agarwalla:2012wf} are extracted, and for atmospheric neutrino experiments which constrain $|\varepsilon_{\mu\tau}^m|\sim {\cal O}(10^{-2}-10^{-3})$ \cite{GonzalezGarcia:2011my,Salvado:2016uqu}. Also reactor as well as long baseline experiments have been probed to be useful, in particular, to restrict the various $|\varepsilon_{e\alpha}^{s,d}|\sim {\cal O}(10^{-2})$ \cite{Biggio:2009nt,Agarwalla:2014bsa} and  $|\varepsilon_{\mu\tau}^m|\sim {\cal O}(0.1)$ \cite{Adamson:2013ovz}, respectively. Although the bounds achieved from non-oscillation experiments \cite{Ohlsson:2012kf}-\cite{Dutta:2020che}  are strong and robust at the level of (generally speaking) percentage, running and planned long baseline experiments aspire to collect large statistics samples which will make possible to reveal feeble effects generated by NSI parameters \cite{Adhikari:2012vc}-\cite{Verma:2019oeb}; in this panorama, the DUNE experiment \cite{Acciarri:2016crz}-\cite{Abi:2020evt} places itself in a relevant position thanks to the capability of improving the bounds on $\varepsilon_{ee, e\mu,e\tau}^{m}$ by $\sim$ 10\% to roughly a factor of 3 \cite{deGouvea:2015ndi}-\cite{Ghoshal:2019pab}. 
However, as discussed in \cite{Blennow:2016etl}, the DUNE Far Detector (FD) is expected to be less performing in constraining source and detector NSIs. Indeed, the bounds obtained in their analysis with $\varepsilon^m=0$ and summarized in Tab. \ref{tabuno}, are just a 10-40\% improvement with respect to the existing literature pertinent to long baseline experiments. The bounds refer only to the moduli of the five parameters $\varepsilon_{\mu e}^s$, $\varepsilon_{\mu\mu}^s$, $\varepsilon_{\mu\tau}^s$, $\varepsilon_{\mu e}^d$ and $\varepsilon_{\tau e}^d$, since the dependence to the other source and detector NSI couplings is only subdominant. Moreover, these constraints are further relaxed when propagation NSI are taken into account into the fit.
\begin{table}[h!]
\begin{center}
\begin{tabular}{|c|c|}
\hline
\textit{\textbf{Parameter}}  & \multicolumn{1}{l|}{\textit{\textbf{DUNE FD 90\% CL bounds}}} \\ \hline
$|\varepsilon_{\mu e}^s|$    & 0.017                                                         \\ \hline
$|\varepsilon_{\mu \mu}^s|$  & 0.070                                                         \\ \hline
$|\varepsilon_{\mu \tau}^s|$ & 0.009                                                         \\ \hline
$|\varepsilon_{\mu e}^d|$    & 0.021                                                         \\ \hline
$|\varepsilon_{\tau e}^d|$   & 0.028                                                         \\ \hline
\end{tabular}
\caption{\label{tabuno} \it 90\% CL limits on the source and detector NSI parameters obtained in \cite{Blennow:2016etl} using the DUNE Far Detector analysis for a total of 10 years of data taking. New phases are unconstrained.}
\end{center}
\end{table}

In this paper we want to (partially) fill the gap, trying to constrain a subset of the $\varepsilon^{s,d}$ matrix elements by means of data that will be collected at the DUNE Near Detector (ND) only. Since the ND is not affected by NSI in the same way as the FD \cite{Blennow:2016jkn}, we expect on the one side to scrutinize more in details those parameters also accessible at the FD and, on the other hand, to access to a complete new set of parameters on which the DUNE FD is not particularly sensitive.  In this context, the role of the ND is promoted as a complementary tool to FD studies \cite{Berryman:2019dme}-\cite{Choubey:2016fpi}, 
more than a mere (although important) indicator of  fluxes and detection cross sections \cite{KayisTopaksu:2019dii}.
Differently from previous works which use the DUNE ND data to probe source and detector NSI parameters \cite{Bakhti:2016gic,Blennow:2016etl}, we provide an analytical discussion of our results. Moreover, we did not consider any assumption on the NSI matrices and we took into account more realistic hypotheses on the systematic uncertainties, including in the analysis also the $\nu_\mu\to\nu_\tau$ oscillation channel, never considered before.

The paper is organized as follows: in Sect.\ref{probs} we derive the approximate transition probabilities relevant for the DUNE ND up to second order in the small entries of the  $\varepsilon^{s,d}$ matrices; in Sect.\ref{sect:per} we discuss in details the performance of the DUNE ND in constraining some of the entries of the above-mentioned matrices while in Sect.\ref{concl} we draw our conclusions. In the Appendix we report our analytical as well as numerical studies of the precision achievable in the measurement of (possible) non-vanishing NSI's.

\section{Transition probabilities at DUNE ND}
\label{probs}
Oscillation probabilities can be obtained from the squared-amplitudes $|\braket{\nu_\beta^d|\nu_\alpha^s}|^2$ which, considering eqs.(\ref{eqs})-(\ref{eqd}), assume the form:
\begin{equation}
   P_{\alpha \beta}= |\braket{\nu_\beta^d|\nu_\alpha^s}|^2=|(1+\varepsilon^d)_{\gamma \beta} (e^{-i H L})_{\gamma\delta} (1+\varepsilon^s)_{\alpha\delta}|^2\,,
\end{equation}
where $L$ is the source-to-detector distance and the Hamiltonian $H$ is given by:
\begin{equation}
    H=\frac{1}{2E}\left[ U \left( \begin{matrix} 0 & 0 & 0 \\ 0 & \Delta m_{21}^2 & 0 \\ 0 & 0 & \Delta m_{31}^2 \end{matrix} \right) U^\dagger+A \left( \begin{matrix} 1+\varepsilon^m_{ee}-\varepsilon^m_{\mu\mu} & \varepsilon^m_{e\mu} & \varepsilon^m_{e\tau} \\ \varepsilon^{m*}_{e \mu} & 0 & \varepsilon^m_{\mu\tau} \\ \varepsilon^{m*}_{e \tau} & \varepsilon^{m*}_{\mu\tau} & \varepsilon^m_{\tau\tau}-\varepsilon^m_{\mu\mu} \end{matrix} \right) \right]\,.
\end{equation}
Here  $U$ is the  PMNS matrix for three active neutrinos and $A$ is the standard matter potential.
Since near detectors are generally placed at distances of ${\cal O}(10^2-10^3)$ meters from the source, the propagation term can be safely neglected for neutrino energies of $\mathcal{O}({\rm GeV})$. Thus, the transition probabilities can be simplified to:
\begin{equation}
   P_{\alpha \beta}= |[(1+\varepsilon^d)^T(1+\varepsilon^s)^T]_{\beta\alpha}|^2.
   \label{probl0}
\end{equation}
Considering that the oscillation phase $\left({\Delta m_{31}^2 L/4E}\right)\sim {\cal O}(10^{-3})$ and the current bounds on $\varepsilon^{s,d}$ are of the order of $10^{-1}-10^{-2}$ \cite{Biggio:2009nt}-\cite{Dutta:2020che}, we expect the approximation in eq.(\ref{probl0}) to be reliable up to the second order in $\varepsilon$.
Parameterizing the new physics complex parameters as $\varepsilon_{\alpha\beta}^{s/d}=|\varepsilon_{\alpha\beta}^{s/d}|e^{i\Phi_{\alpha\beta}^{s/d}}$, the disappearance probabilities ($\alpha = \beta$) read:
\begin{equation}
\begin{split}
    P_{\alpha\alpha}=&1+2|\varepsilon_{\alpha\alpha}^s|\cos{\Phi_{\alpha\alpha}^s}+2|\varepsilon_{\alpha\alpha}^d|\cos{\Phi_{\alpha\alpha}^d}+|\varepsilon_{\alpha\alpha}^s|^2+|\varepsilon_{\alpha\alpha}^d|^2+ \\
&+4|\varepsilon_{\alpha\alpha}^s||\varepsilon_{\alpha\alpha}^d|\cos{\Phi_{\alpha\alpha}^s}\cos{\Phi_{\alpha\alpha}^d}+ 2\sum_{\beta \neq \alpha} |\varepsilon_{\alpha\beta}^s||\varepsilon_{\beta\alpha}^d|\cos{(\Phi_{\alpha\beta}^s+\Phi_{\beta\alpha}^d)}\,,
\end{split}
\end{equation}
while the appearance probabilities ($\alpha \ne \beta$) are given by:
\begin{equation}
    P_{\alpha\beta}=|\varepsilon_{\alpha\beta}^s|^2+|\varepsilon_{\alpha\beta}^d|^2+2|\varepsilon_{\alpha\beta}^s||\varepsilon_{\alpha\beta}^d|\cos{(\Phi_{\alpha\beta}^s-\Phi_{\alpha\beta}^d)}.
\label{app_for}
\end{equation}
In the disappearance case, the dependence on the diagonal NSI parameters appears already at the first order and the whole probabilities (including second-order corrections driven by the off-diagonal matrix elements) depend on twelve independent real parameters; in addition, the leading order and the diagonal next-to-leading terms  display a complete symmetry under the interchange $s \leftrightarrow d$, so that we expect similar sensitivities to $\varepsilon^{s,d}_{\alpha\alpha}$. The off-diagonal second order corrections are no longer symmetric since two flavor changes are needed to have the same flavor at the source and at the detector.

In the appearance case, the new parameters appear  at the second order and only four independent of them  are involved. The relevant $P_{\mu e}$ and $P_{\mu \tau}$ are completely symmetric under $s \leftrightarrow d$  because, at short distances, the flavor changing can happen at both  source or detector with no fundamental distinction. 

The drastic reduction of independent NSI parameters the ND is sensitive to, allows to derive simple rules on how their admitted ranges can be strongly limited compared to the existing literature.
Indeed, let us work in the simplified scenario where the experiment counts a certain number $N$ of events when searching for $\nu_\alpha \to \nu_\beta$ oscillations; since the probabilities in eq.(\ref{probl0}) show no dependence on neutrino energy, baseline, matter potential and standard mixing parameters,
$N$ assumes the form:
\begin{equation}
    N= N_0 P_{\alpha\beta}(\varepsilon^s,\varepsilon^d)\,,
\end{equation}
where the normalization factor $N_0$ includes all the detector properties and, given an observation mode $\nu_\alpha\to\nu_\beta$, is defined by:
\begin{eqnarray}
\label{eq:asirate2}
N_0^{\alpha\beta} &=&\int_{E_\nu} dE_\nu \,\sigma_\beta(E_\nu)\,\frac{d\phi_\alpha}{dE_\nu}(E_\nu) \,\varepsilon_\beta(E_\nu)\,,
\end{eqnarray} 
in which $\sigma_\beta$ is the the cross section for producing the lepton $\beta$, $\varepsilon_\beta$ the detector efficiency and $\phi_\alpha$ the initial neutrino flux of flavor $\alpha$.
Suppose now that we want to exclude a region of the parameter space using a simple $\chi^2$ function defined as:
\begin{equation}
    \chi^2=\frac{(N_{obs}-N_{fit})^2}{\sigma^2}\,,
\end{equation}
where $\sigma$ represents the statistical uncertainty on the number of events. Assuming vanishing true values of all NSI parameters,  the $\chi^2$ function becomes:
\begin{equation}
    \chi^2=\frac{N_0^2}{\sigma^2}[\delta_{\alpha\beta}- P_{\alpha\beta}(\varepsilon_{fit}^s, \varepsilon_{fit}^d)]^2\,.
\label{chi}
\end{equation}
For appearance analysis,  eq.(\ref{app_for}) allows us to write:
\begin{equation}
    \chi^2=\frac{N_0^2}{\sigma^2}[|\varepsilon_{\alpha\beta}^s|^2+|\varepsilon_{\alpha\beta}^d|^2+2|\varepsilon_{\alpha\beta}^s||\varepsilon_{\alpha\beta}^d|\cos{(\Phi_{\alpha\beta}^s-\Phi_{\alpha\beta}^d})]^2\,,
\end{equation}
whose minimum can always be found when $\cos{\Delta \Phi}=-1$. Thus, for every pairs of $(|\varepsilon_{\alpha\beta}^s|,|\varepsilon_{\alpha\beta}^d|)$:
\begin{equation}
\label{chiminapp}
    \chi_{min}^2=\frac{N_0^2}{\sigma^2}(|\varepsilon_{\alpha\beta}^s|-|\varepsilon_{\alpha\beta}^d|)^4\,.
\end{equation}
Indicating with $\chi^2_{0,\alpha\beta}$ the value corresponding to the cut of the $\chi^2$ at a given CL,
we can exclude the region delimited by:
\begin{equation}
\label{app}
   ||\varepsilon_{\alpha\beta}^s|-|\varepsilon_{\alpha\beta}^d||
>\sqrt[4]{\frac{ \chi^2_{0,\alpha\beta}\, \sigma^2}{N_0^2}}\,,
\end{equation}
which is external to a band in the $(|\varepsilon_{\alpha\beta}^s|,|\varepsilon_{\alpha\beta}^d|)$-plane of width

\begin{equation}\label{Deltaalphabeta}
    \Delta_{\alpha\beta} = \sqrt[4]{\frac{ 4\chi^2_{0,\alpha\beta}\, \sigma^2}{N_0^2}}
\end{equation}
centered on the line $|\varepsilon_{\alpha\beta}^s|=|\varepsilon_{\alpha\beta}^d|$. Thus, $\Delta_{\alpha\beta}$ provide a measure of the allowed parameter space.
Clearly, the excluded region is larger when the uncertainty on the number of events $\sigma$ is smaller and the normalization factor $N_0$ is bigger.\\
Consider now the disappearance case; neglecting second order terms, the $\chi^2$ function is now:
\begin{equation}
    \chi^2=\frac{4N_0^2}{\sigma^2}(|\varepsilon_{\alpha\alpha}^s|\cos{\Phi_{\alpha\alpha}^s}+|\varepsilon_{\alpha\alpha}^d|\cos{\Phi_{\alpha\alpha}^d})^2=\frac{4N_0^2}{\sigma^2}[\Re(\varepsilon_{\alpha\alpha}^s)+\Re(\varepsilon_{\alpha\alpha}^d)]^2\,.
\end{equation}
Following the same procedure as for the appearance case, the excluded region in the $[\Re(\varepsilon_{\alpha\alpha}^s), \Re(\varepsilon_{\alpha\alpha}^d)]$-plane is delimited by: 
\begin{equation}
\label{dis}
   |\Re(\varepsilon_{\alpha\alpha}^s)+\Re(\varepsilon_{\alpha\alpha}^d)|
>\sqrt{\frac{ \chi^2_{0,\alpha\alpha} \,\sigma^2}{4N_0^2}}\,,
\end{equation}
where, in this case, the band width is: 
\begin{equation}\label{Deltaalphaalpha}
\Delta_{\alpha\alpha}=\sqrt{\frac{ \chi^2_{0,\alpha\alpha} \,\sigma^2}{2N_0^2}}
\end{equation}
with $\chi^2_{0,\alpha\alpha}$ being the desired cut of the $\chi^2$. Notice that, for the same $\sigma$ and $N_0$, we expect the disappearance channels alone to be  more performing than the appearance ones. This is essentially motivated by the absence of first order terms in $\varepsilon$ in the appearance probabilities.
Notice also that eqs.(\ref{app}) and (\ref{dis}) show a perfect symmetry under the interchange of  source and detector parameters which, however, could be (partially) disentangled  if a multi-channel analysis is performed. For example, the parameter $|\varepsilon_{\mu e}^s|$ appears in the $\nu_\mu \to \nu_e$ oscillation but also as a correction to the $\nu_\mu \to \nu_\mu$ probability, differently from the case of $|\varepsilon_{\mu e}^d|$ which is present in the $\mu \to e$ transition only. Nevertheless, given the relatively small contributions of the second order terms compared to the first order, we expect such corrections to have a negligible impact.  

\section{Performance of the DUNE Near Detector}
\label{sect:per}
In this section we will provide the details  of our numerical simulation and present the sensitivity of the DUNE ND to the NSI parameters discussed above. 
\subsection{The DUNE Near Detector}

DUNE (Deep Underground Neutrino Experiment) is one of the most promising future neutrino oscillation experiments. It will be situated in the USA, where a $\nu_\mu$ beam from FNAL will be focused to SURF (Sanford Underground Research Facility), 1300 km away, where the Far Detector complex is under construction \cite{Acciarri:2016crz}-\cite{Abi:2020evt}. 
Recent studies have contemplated the possibility of three modules for the DUNE ND \cite{ND1}-\cite{ND5}: 
\begin{itemize}
    \item A Liquid Argon Time Projection Chamber (TPC) situated at 574 m from the neutrino source (ArgonCube). The main purposes of this detector are flux and cross section measurements. Its performances in terms of detection efficiencies and systematics can be considered the same as the far detector ones. The total volume of the TPC is 105 $m^3$, while the argon fiducial mass can be considered 50 tons. The PRISM system will be able to move this detector to different off-axis positions in order to have a better determination of the neutrino flux at different angles.
    \item A so called {\it Multi Purpose Detector} (MPD), namely a magnetic spectrometer with a 1 ton High-Pressure Gaseous Argon TPC arranged in the middle of the three module system, whose major field of application will be the study of possible new physics signals.
    \item The System for On-Axis Neutrino Detection (SAND), which will measure the on-axis neutrino flux when ArgonCube is moved to different positions. It will consist of the former KLOE magnet and calorimeter supplemented by a tracker for the escaping particles. 
\end{itemize}
For the following study,  we can neglect the contribution of the MPD and the SAND in our numerical simulations because, due to their limited mass, they would not be able to collect a significant statistics. 
For the simulation of the Liquid Argon TPC Near Detector, we follow the  suggestion of the DUNE collaboration and consider the same configuration as the Far Detector. 
In particular, in order to perform a $\chi^2$ statistical analysis (based on the pull method \cite{Fogli:2002pt}), we used the GLoBES software \cite{Huber:2004ka,Huber:2007ji} supplemented by the NSI package developed in \cite{Kopp:2007ne,Kopp:2006wp}, for which DUNE simulation files have been provided \cite{fluxes,Alion:2016uaj}.

The detection channels considered in the simulations are:
\begin{itemize}
    \item $\nu_\mu$ CC channel, which is composed by events from $\nu_\mu \to \nu_\mu$ oscillations. Background to this channel are misidentified $\nu_\tau$ CC and NC events. 
    \item $\nu_e$ CC channel, which is composed by the $\nu_\mu \to \nu_e$ events (driven by NSI) and by $\nu_e \to \nu_e$ events from the $\nu_e$ beam contamination. Backgrounds for this channel are misidentified $\nu_\mu$ CC, $\nu_\tau$ CC and NC events. 
    \item $\nu_\tau$ CC channel, which is composed by $\nu_\mu \to \nu_\tau$ events driven by NSI. As in \cite{deGouvea:2019ozk,Ghoshal:2019pab}, we have considered events coming from electronic and hadronic $\tau$ decays. Background to this channel are misidentified $\nu_e$ CC and NC events.
\end{itemize}

The systematics considered in our study will be dominated by cross sections and flux normalization uncertainties. While the former could be in principle improved by future data and calculations, the latter will anyway remain as the dominant source of error because of the hadroproduction processes and  uncertainties in the focusing system at the LBNF beam. Differently from similar studies involving the DUNE ND \cite{Ballett:2019xoj,Choubey:2016fpi}\footnote{In these papers different physics models than NSI are analyzed, less sensitive to  systematics.}, where the same systematic uncertainties reported in the DUNE Far Detector GLoBES configuration file have been used, we decided to consider worst systematics since the ND cannot benefit of a (partial) systematic cancellation provided by a detector closer to the neutrino production region. In particular, we took into account an overall systematic normalization uncertainty of 10\% for the $\nu_\mu$ disappearance, $\nu_e$ disappearance and $\nu_e$ appearance channels signals and of 25\% for the $\nu_\tau$ appearance signal. For the NC background we considered a 15\% uncertainty. Other scenarios with more aggressive and more conservative choices will be studied in Sect. \ref{changesys}.
For all channels, smearing matrices and efficiencies have been taken from \cite{Alion:2016uaj}; for all of them we considered an energy bin width of 125 MeV.
Notice that, since the NSI couplings only change the total number of events in every channel, the energy binning we adopted in our numerical simulations are not strictly necessary. However we prefer to take them into account in order to adhere to the experimental setup proposed by the DUNE collaboration itself. 
\subsection{Simulation Results}
In our numerical simulations we use exact transition probabilities and we set all NSI true values to zero; we marginalize over all absolute values of the  parameters appearing in the probabilities up to the second order (with no priors) and over all relevant phases, which are allowed to vary in the $[0,2\pi)$ range \footnote{The standard oscillation parameters are fixed to the central values reported in \cite{Esteban:2018azc} because they have no effects in our fit.}. 
Since, as showed in the previous section, the strongest constraints on $\varepsilon^{s,d}_{\alpha\beta}$ can be obtained from the corresponding oscillation probability $P_{\alpha\beta}$, we simulated one transition channel at a time.
To make a comparison with the bounds obtainable at the FD (see Tab.(\ref{tabuno})) we  consider $5+5$ years of data taking.

\begin{figure}[ht]
\begin{center}
\includegraphics[height=7cm,width=7cm]{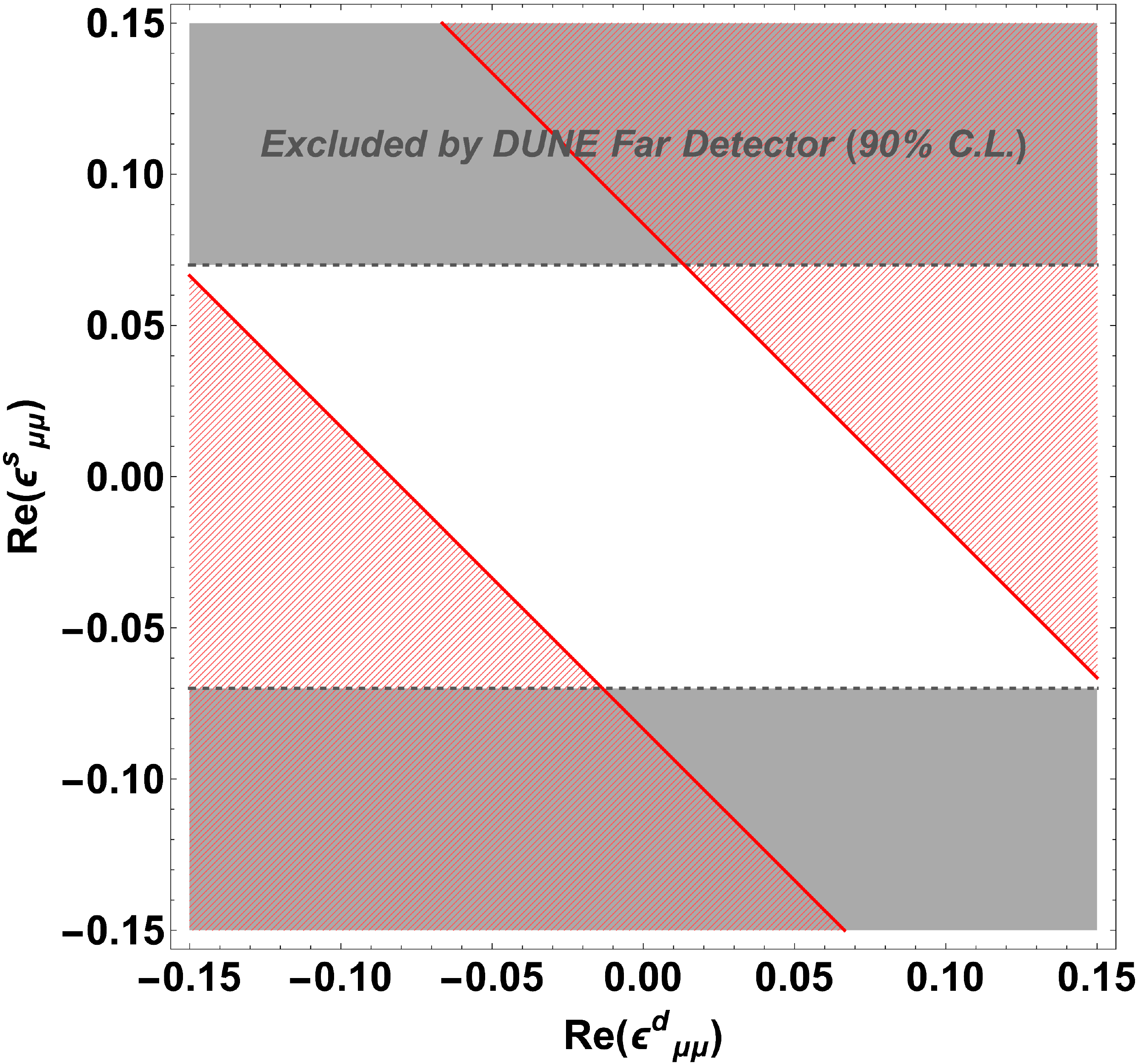}
\includegraphics[height=7cm,width=7cm]{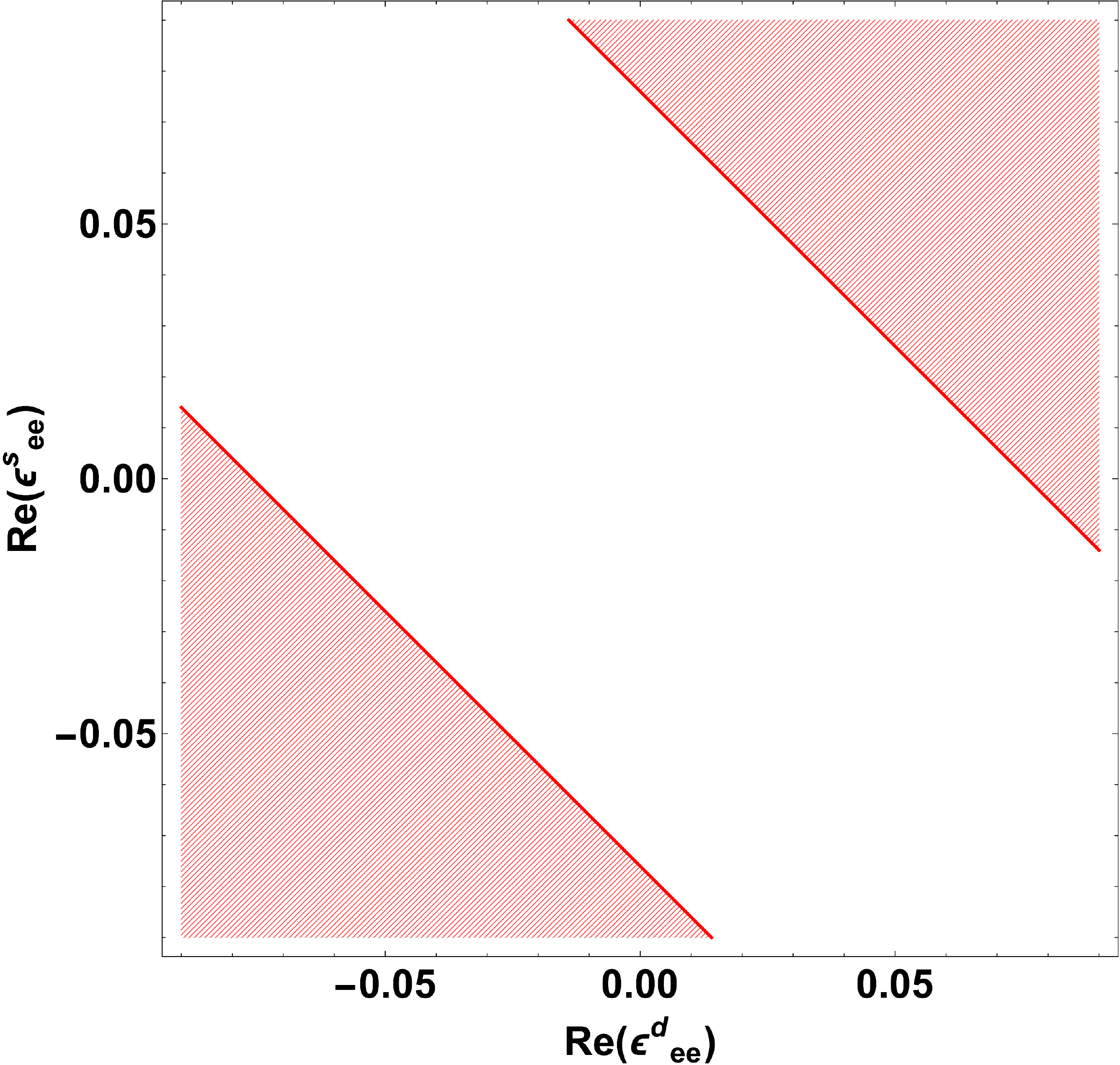}
\caption{\it 90\% CL excluded regions (in red) in the  $\left[\Re(\varepsilon_{\mu\mu}^s),\Re(\varepsilon_{\mu\mu}^d)\right]$-plane (left panel) and $\left[\Re(\varepsilon_{e e}^s),\Re(\varepsilon_{e e}^d)\right]$-plane (right panel) by the DUNE ND. The FD  excluded zones are shown with horizontal gray bands.} 
\label{mumuandee}
\end{center}
\end{figure}

In the disappearance sector, the interesting pairs of NSI parameters for the ND are  $\left[\Re(\varepsilon_{\mu\mu}^s),\Re(\varepsilon_{\mu\mu}^d)\right]$ and $\left[\Re(\varepsilon_{e e}^s),\Re(\varepsilon_{e e}^d)\right]$, which are mainly constrained by the $\nu_\mu \to \nu_\mu$ and $\nu_e \to \nu_e$ transitions, respectively. The regions that could be excluded by the DUNE ND are displayed in Fig.(\ref{mumuandee}), where we also superimposed the limits set by the FD analysis only\footnote{Horizontal and and vertical lines showed in our plots do not represent the results of a correlation analysis at the far detector, but only the sensitivity limits obtained after a full marginalization on the parameters space.} \cite{Blennow:2016etl} (no limits can be put on $\varepsilon_{e e}^{s,d}$).
As it is clear from the left panel, the numerical results completely reflect the analytic anticorrelations discussed in eq.(\ref{dis}): 
even though for every value of $\Re(\varepsilon_{\mu\mu}^d)$ there is an interval of $\Re(\varepsilon_{\mu\mu}^s)$ for which the $\chi^2$ is small, it is nonetheless possible to exclude a sizable portion of the parameter space allowed by FD analysis. Similar considerations can be done on the parameters $\Re(\varepsilon_{e e}^{s,d})$ shown in the right panel, for which the ND is able to rule out a relevant fraction of them, a goal otherwise not possible with the DUNE FD alone.

\begin{figure}[ht]
\begin{center}
\includegraphics[height=7cm,width=7cm]{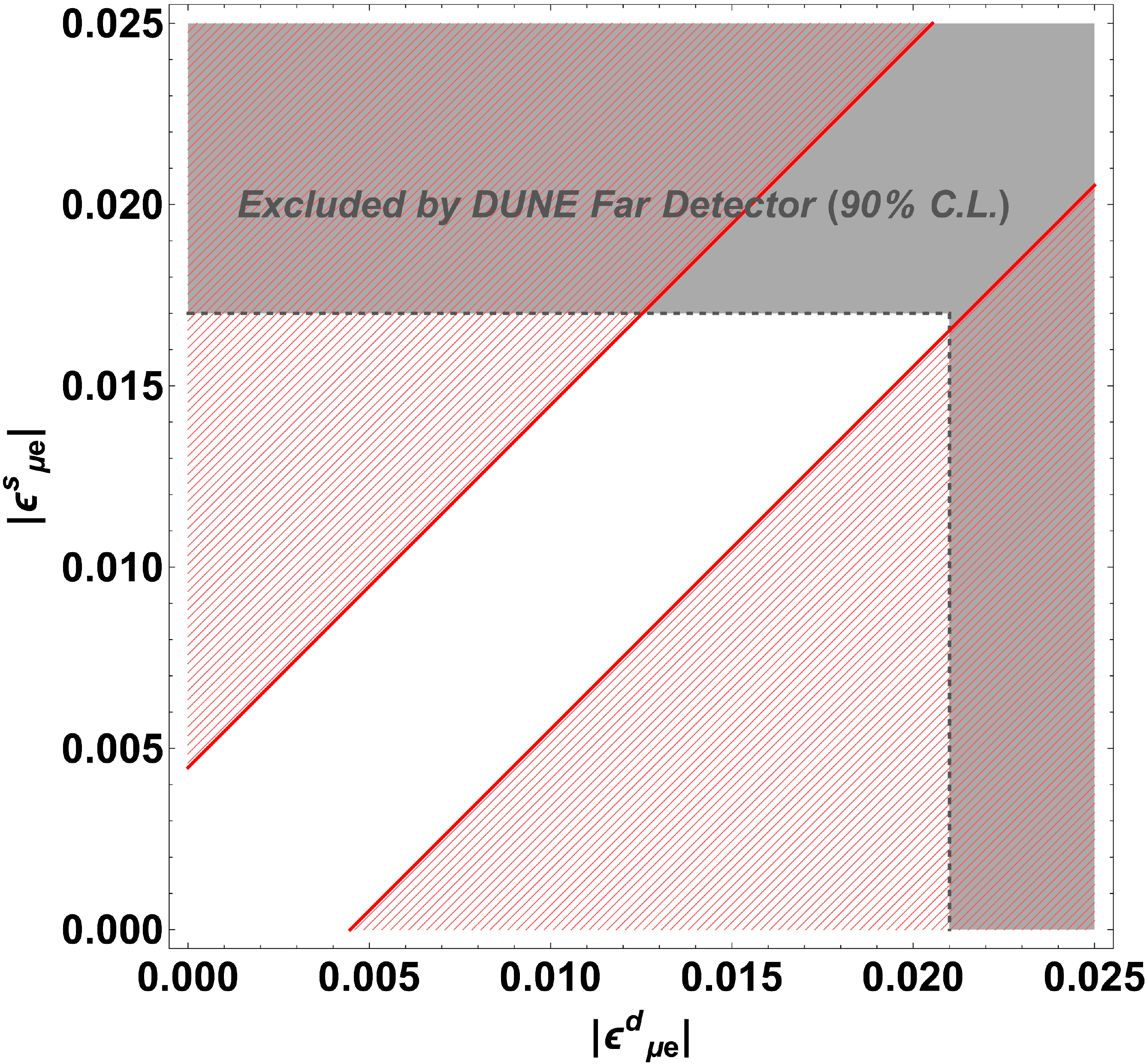}
\includegraphics[height=7cm,width=7cm]{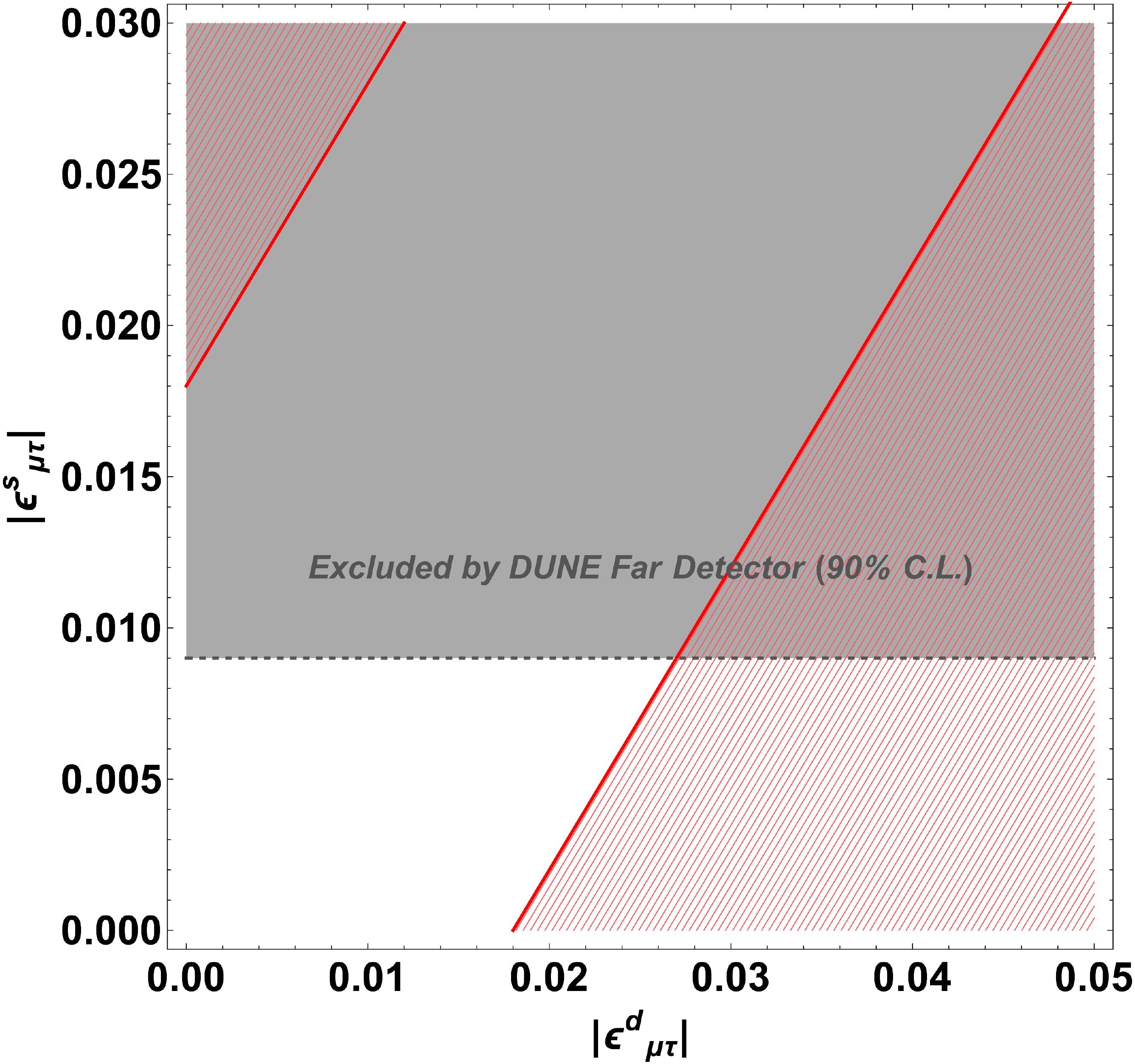}
\caption{\it Same as Fig.(\ref{mumuandee}) but in the $\left(|\varepsilon_{\mu e}^s|,|\varepsilon_{\mu e}^d|\right)$ (left panel) and $\left(|\varepsilon_{\mu \tau}^s|,|\varepsilon_{\mu \tau}^d|\right)$ planes (right panel).} 
\label{mueandmutau}
\end{center}
\end{figure}

\begin{figure}[h!]
\begin{center}
\includegraphics[height=7cm,width=7cm]{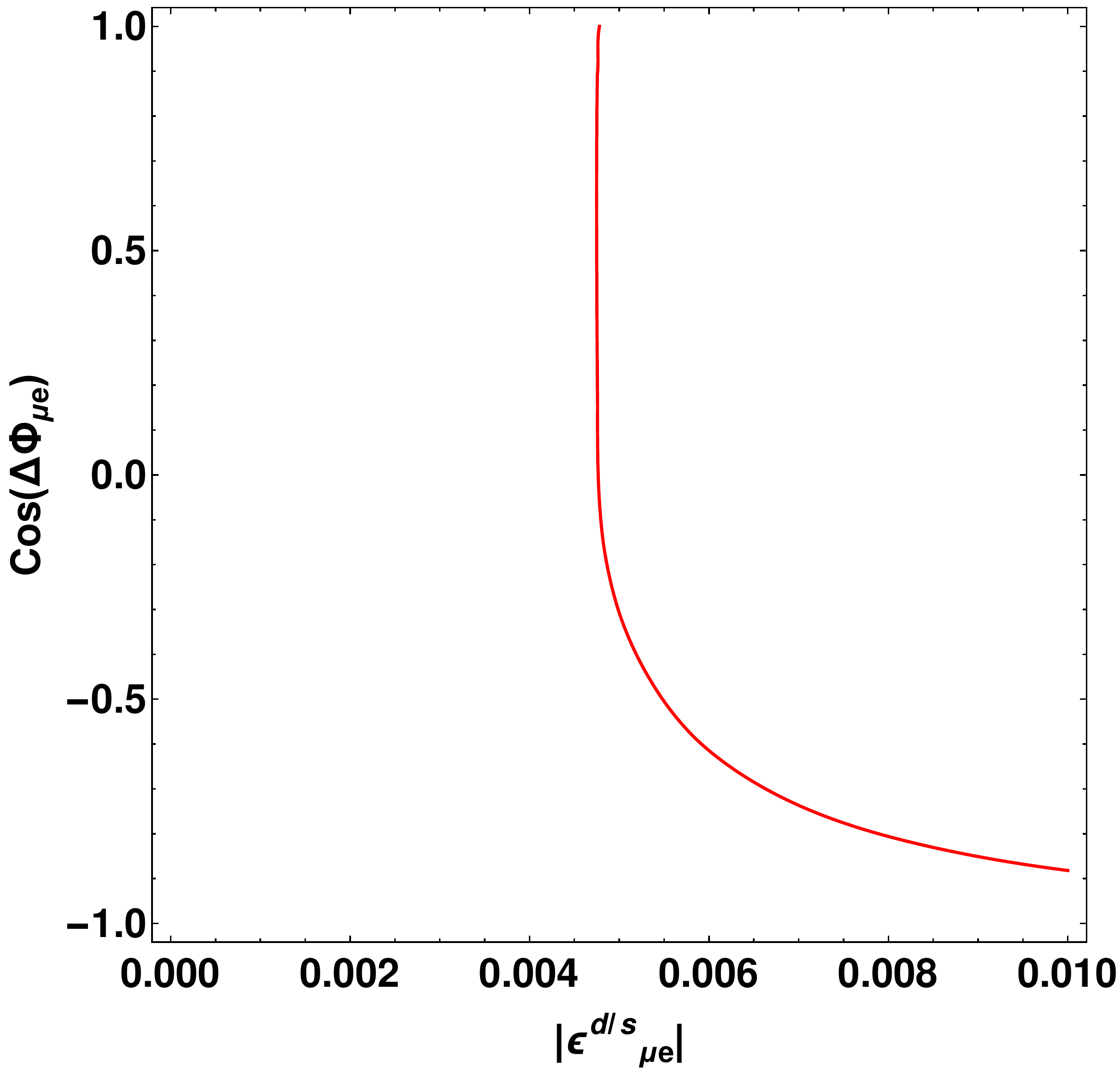}
\includegraphics[height=7cm,width=7cm]{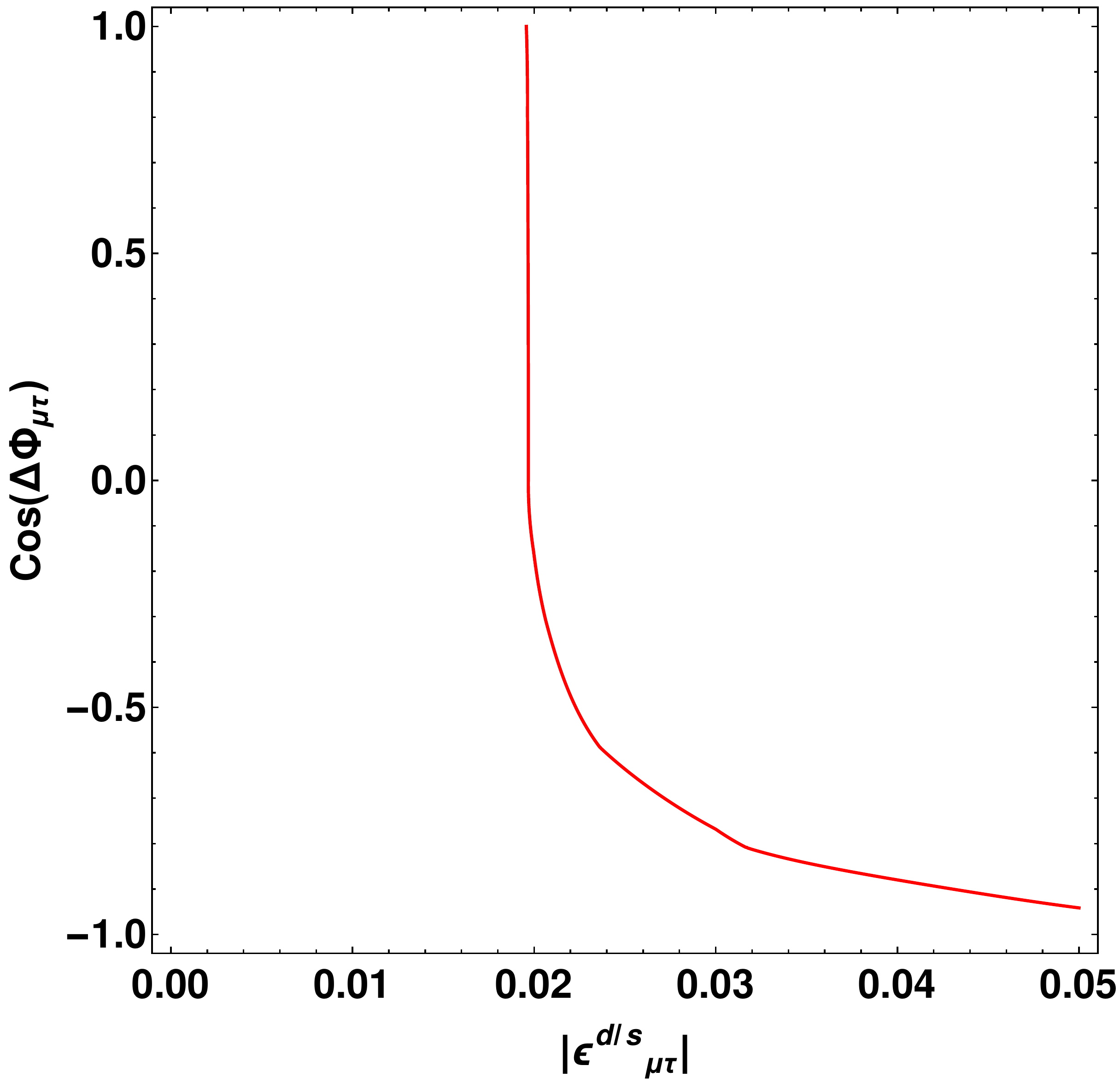}
\caption{\it Contours at 90\% CL in the $\left(|\varepsilon_{\mu e}^{s/d}|, \cos{\Delta \Phi_{\mu e}}\right)$ (left panel) and $\left(|\varepsilon_{\mu \tau}^{s/d}|, \cos{\Delta \Phi_{\mu \tau}}\right)$ (right panel) planes obtained  by our DUNE ND simulations.} 
\label{corr}
\end{center}
\end{figure}

Given the band width in eq. (\ref{Deltaalphaalpha}), the above considerations can be summarized as follows: 
\begin{equation}
\label{bandsdis}
\Delta _{\mu\mu} = 0.12 \qquad \Delta _{ee} = 0.11\;.
\end{equation}
Notice that for 90\% CL $\chi^2_{0}= 4.6$ (2 degrees of freedom) and $N_0$ is roughly $10^7$ events per year for the $\nu_\mu\to\nu_\mu$ channel and $10^5$ events per year in the $\nu_e\to\nu_e$ channel.
The obtained values of the band are of the same order of the systematics discussed in the previous section, namely 10\% for the signal and 15\% for the background, and are almost the same for the two channels even though the number of $\nu_\mu$ events is two order of magnitude larger than the number of $\nu_e$ events. This reflects the fact that, for the disappearance channels, we cannot be sensitive to NSI parameters which cause changes to standard oscillation probabilities smaller than the adopted systematic uncertainties (for further discussions about systematics see Sec. \ref{changesys}).
However, even with our realistic assumptions, the result on $\Delta_{\mu\mu}$ permits to exclude parts of the parameter space allowed  by the general analysis performed in \cite{Biggio:2009nt} ($|\varepsilon_{\mu\mu}^s|<0.068$ and $|\varepsilon_{\mu\mu}^s|<0.078$). On the other hand, the result on $\Delta_{ee}$ is worse than the one set by reactor experiments like Daya Bay \cite{Agarwalla:2014bsa} ($|\varepsilon_{ee}|<2\times10^{-3}$) obtained, we have to outline, under the restrictive assumption $\varepsilon^s=\varepsilon^{d*}$.
In the case of the appearance channels, eq.(\ref{app_for}) highlights that the interesting pairs of parameters  are $\left(|\varepsilon_{\mu e}^s|,|\varepsilon_{\mu e}^d|\right)$ and $\left(|\varepsilon_{\mu \tau}^s|,|\varepsilon_{\mu \tau}^d|\right)$. We show the 90\% CL excluded regions in Fig.(\ref{mueandmutau}) where we also displayed the bounds that would be set by the FD.
Also in this case, the correlations outlined in eq.(\ref{app}) is recovered and large portions of the parameter spaces  can be ruled out, in particular for the $\nu_e$ appearance channel. On the other hand, the small signal to background ratio in the $\nu_\tau$ appearance channel results in a larger band width; thus, the region excluded by the ND but allowed by the FD is delimited by $|\varepsilon_{\mu\tau}^d|>0.024$. The appearance results are summarized by the following widths:
\begin{equation}
   \Delta_{\mu e} = 0.0065 \qquad \Delta_{\mu \tau} =0.026\,.
\label{appwidth}
\end{equation}
A better background rejection in the $\nu_\tau$ channel could reduce the band width by up to one order of magnitude.
An important role in defining the allowed ranges for the appearance parameters $|\varepsilon_{\mu e}^{s/d}|$ and $|\varepsilon_{\mu \tau}^{s/d}|$ is played by the CP violating phases $\Phi_{\alpha\beta}^s$ and $\Phi_{\alpha\beta}^d$. Recalling eq.(\ref{chiminapp}), it is clear that the degeneracy that let the $\chi^2$ vanish when the absolute values of detector and source parameters are the same, occurs only when $\Delta \Phi_{\alpha\beta}=\Phi_{\alpha\beta}^s-\Phi_{\alpha\beta}^d$ is very close to $\pi$. For all other values of the phase difference, the ND could be able to set very stringent 90\% CL limits (with a 5+5 years of data taking), namely: 
\begin{equation}
|\varepsilon_{\mu e}^{s/d}|<0.0046 \qquad  |\varepsilon_{\mu \tau}^{s/d}|<0.019\,, 
\end{equation}
which are very competitive to the ones set so far by other neutrino oscillation experiments (for instance, $|\varepsilon_{\mu e, \mu\tau}^{s/d}|<\mathcal{O}(10^{-2})$ obtained in \cite{Biggio:2009nt} and \cite{Blennow:2015nxa}).
This is clearly shown in Fig.(\ref{corr}) where we present  the contours at 90\% CL in the $\left(|\varepsilon_{\mu e}^{s/d}|, \cos{\Delta \Phi_{\mu e}}\right)$ and $\left(|\varepsilon_{\mu \tau}^{s/d}|, \cos{\Delta \Phi_{\mu \tau}}\right)$-planes, obtained after marginalizing the $\chi^2$ function over all undisplayed parameters.

\subsection{Changing systematic errors}
\label{changesys}

As discussed in the previous sections, the choice of the systematic uncertainties is a crucial point in the determination of the limits that the Near Detector could be able to set. In order to understand how much the band widths $\Delta_{\alpha\beta}$ would change in the case of a different choice of systematics, we performed the same simulations for a data taking time of 5+5 years considering three different cases:
\begin{itemize}
    \item \textit{Case A}: the standard (optimistic) case, namely the one implemented in the DUNE Far Detector GLoBES configuration file. In this case the systematics are 5\% for the $\nu_\mu$ disappearance channel, 2\% for the $\nu_e$ appearance and disappearance channels and 20\% for the $\nu_\tau$ appearance channel. The uncertainty on the NC background has been considered to be 10\%.
    \item \textit{Case B}: the more realistic choice used in the previous section, where we fixed 10\% for the $\nu_e$ appearance, $\nu_e$ disappearance and $\nu_\mu$ disappearance, 25\% for the $\nu_\tau$ appearance and 15\% for the NC background.
    \item \textit{Case C}: a more pessimistic case in which the systematics are 15\% for $\nu_e$ appearance, $\nu_e$ disappearance and $\nu_\mu$ disappearance, 30\% for the $\nu_\tau$ appearance and 20\% for the NC background.
\end{itemize}

\begin{figure}[ht]
\begin{center}
\includegraphics[height=7cm,width=8cm]{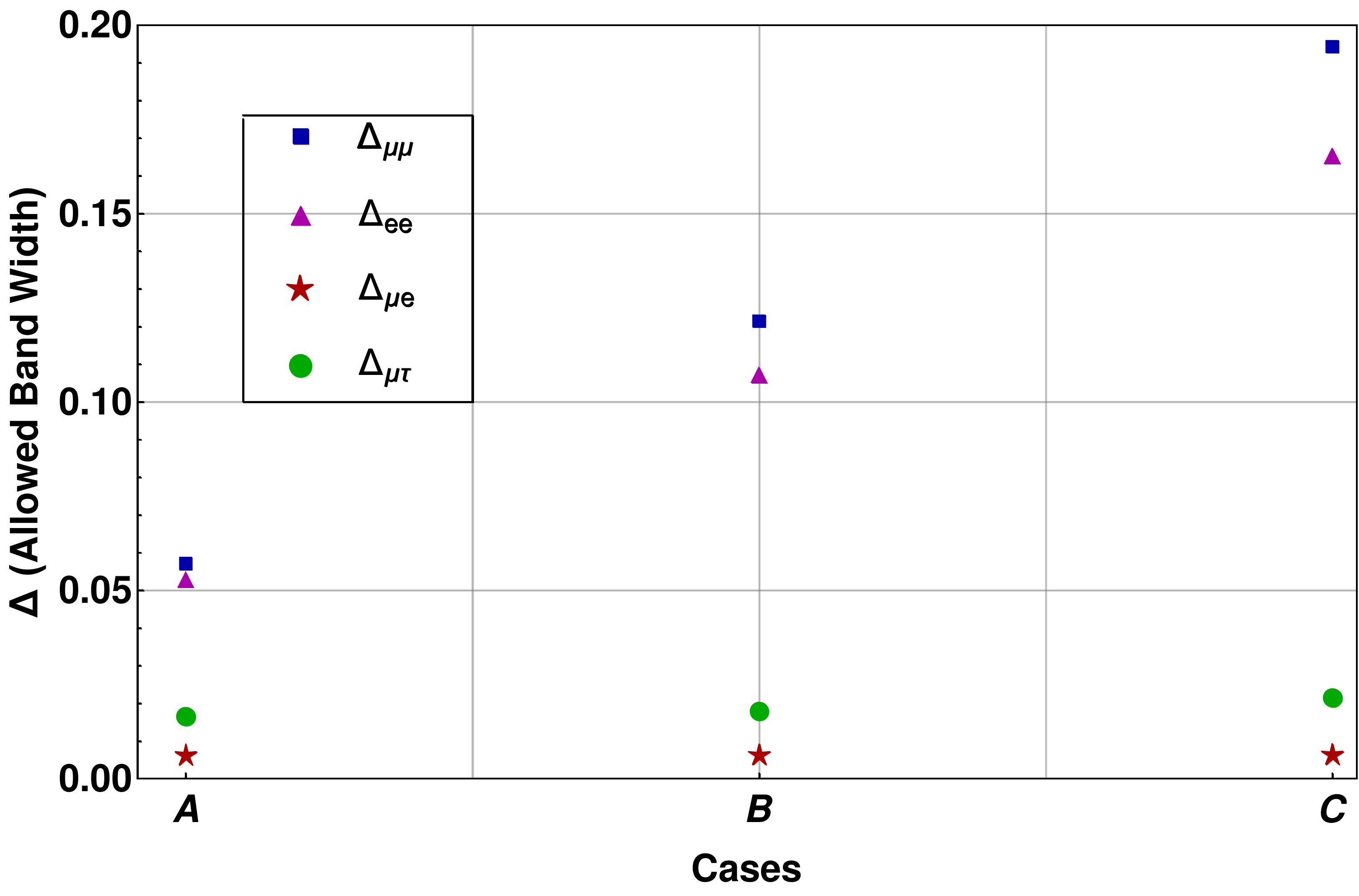}
\caption{\it  Variation of the allowed band widths $\Delta_{\alpha\beta}$ for a data taking time of 5+5 years and four different choice of systematics: optimistic (A), standard (B) and pessimistic (C). } 
\label{sys}
\end{center}
\end{figure}
The results of our simulations are reported in Fig.\ref{sys}. We  clearly see that $\Delta_{ee}$ and $\Delta_{\mu\mu}$ are the parameters which are affected the most by the systematics, as previously discussed. Indeed, being the survival probability at L=0 in the standard model equal to 1, the number of observed events will be, even in presence of the small effect of the NSI, of the same order of magnitude as $N_0$. Thus, when statistical errors are negligible, the definition for the band width (eq.(\ref{Deltaalphaalpha})) can be simplified to: 
\begin{equation}
   \Delta_{\alpha\alpha}\sim\sqrt{\frac{\chi^2_{0,\alpha\alpha}}{2}}\sigma_{sys}\,,
\end{equation}
where we used $\sigma\sim N_0\sigma_{sys}$
.\\
For the appearance parameters, we register a less evident increasing of the band widths passing from the case (A) to (C), since in this case statistic uncertainties are always dominating over systematics. Indeed, for the two appearance channels, $N_0$ is $\sim 10^7$ per year in the $\nu_\mu\to\nu_e$ channel and $\sim 10^6$ per year in the $\nu_\mu\to\nu_\tau$ channel, but the observed number of events is small due to the very short baseline. For a given small number of observed events $N_{obs}$, we have  $\sigma^2=N_{obs}^2\sigma^2_{sys}+N_{obs}\sim N_{obs}$. Thus, the width can be simplified as follows: 
\begin{equation}   \Delta_{\alpha\beta}\sim\sqrt[4]{\frac{4\chi^2_{0,\alpha\beta}N_{obs}}{N_0^2}}\,.
\end{equation}
This quantity is roughly independent on the systematics and for $N_{obs}=\mathcal{O}(10)$ is of the order of $10^{-3}$. This number is in agreement with our numerical results for $\Delta_{\mu e}$ while for $\Delta_{\mu\tau}$ the agreement is confined to the case where the NC background (which in addition suffers by the increase of the systematics) is turned off.

We want to outline that  we recomputed the various $\Delta_{\alpha\beta}$ also
for several positioning of the DUNE ND at different off-axis angles with respect to the beam direction \cite{fluxes} and found a general worsening of the ND performances due to the decreased number of collected events. In fact, spectra distortions of signal and backgrounds cannot improve source and detector NSI analysis since probabilities in this regime do not depend on neutrino energies.

\section{Conclusions}
\label{concl}
In this paper we have discussed in details the role of the DUNE Near Detector in constraining some of the source and detector NSI parameters. We have derived useful analytic expressions for the appearance and disappearance transition probabilities at zero-baseline, up to the second order in the small $\varepsilon_{\alpha\beta}^{s,d}$. 
We have shown that the allowed regions in the planes $\left[\Re(\varepsilon_{\mu\mu}^s),\Re(\varepsilon_{\mu\mu}^d)\right]$, $\left[\Re(\varepsilon_{e e}^s),\Re(\varepsilon_{e e}^d)\right]$, $\left(|\varepsilon_{\mu e}^s|,|\varepsilon_{\mu e}^d|\right)$   and $\left(|\varepsilon_{\mu \tau}^s|,|\varepsilon_{\mu \tau}^d|\right)$ follow the shapes identified by our analytic considerations and result strongly constrained if compared to the DUNE Far Detector studies, when available. Furthermore, restrictive bounds can be set with a total of $10$ years on $\nu + \bar \nu$ data taking:  
$|\varepsilon_{\mu e}^{s/d}|<0.0045$ and $|\varepsilon_{\mu \tau}^{s/d}|<0.019$, for a wide range of the values of the related phases.
Finally, we showed that if we increase the systematic uncertainties (which are a crucial ingredient at the Near Detector because flux and cross section uncertainties cannot be easily reduced), the two band widths which show a significant worsening are $\Delta_{\mu\mu}$ and $\Delta_{ee}$. The other two band widths suffer by only a small increase in amplitude and the related parameter spaces remain drastically reduced with respect to the current ranges allowed by oscillation experiments.

\section*{Appendix: On the achievable precision on non-vanishing NSI's}

The relatively simple strategy we used to find analytic bounds on NSI parameters can also be applied to compute the precision on the measurement of non-vanishing parameters, that is in the case where the true values of the source and detector parameters are non zero. In this case eq.(\ref{chi}) becomes:
\begin{equation}
    \chi^2=\frac{N_0^2}{\sigma^2}[\delta_{\alpha\beta}+K_{\alpha\beta} -P_{\alpha\beta}(\varepsilon_{fit}^s,\varepsilon_{fit}^d)]^2\,,
\end{equation}
where $K_{\alpha\beta}$ is defined as the true $P_{\alpha\beta}$ for the appearance channels and $P_{\alpha\alpha}-1$ for the disappearance channels.

Let us start from the disappearance. Given the structure of the $\chi^2$ function:
\begin{equation}
    \chi^2=\frac{4N_0^2}{\sigma^2}\left[ K_{\alpha\alpha}/2-\Re(\varepsilon_{\alpha\alpha}^s)-\Re(\varepsilon_{\alpha\alpha}^d) \right] ^2\,,
\end{equation}
the allowed regions in the 
$\left[\Re(\varepsilon_{\alpha\alpha}^s),\Re(\varepsilon_{\alpha\alpha}^d)\right]$-plane are identified by:
\begin{equation}
\label{disallowed}
\left| \Re(\varepsilon_{\alpha\alpha}^d)+\Re(\varepsilon_{\alpha\alpha}^s)-K_{\alpha\alpha}/2\right|< \sqrt{\frac{ \chi_{0,\alpha\alpha}^2 \sigma^2}{4N_0^2}}  \,.
\end{equation}
This means that the allowed regions around the values of $\Re(\varepsilon_{\alpha\alpha}^d)$ and $\Re(\varepsilon_{\alpha\alpha}^s)$ chosen by Nature have essentially similar shapes as those presented in Fig.(\ref{mumuandee}) but with a band centered on the line $\Re(\varepsilon_{\alpha\alpha}^d)=-\Re(\varepsilon_{\alpha\alpha}^s)+K_{\alpha\alpha}/2$.

\begin{figure}[h!]
\begin{center}
\includegraphics[height=7cm,width=7cm]{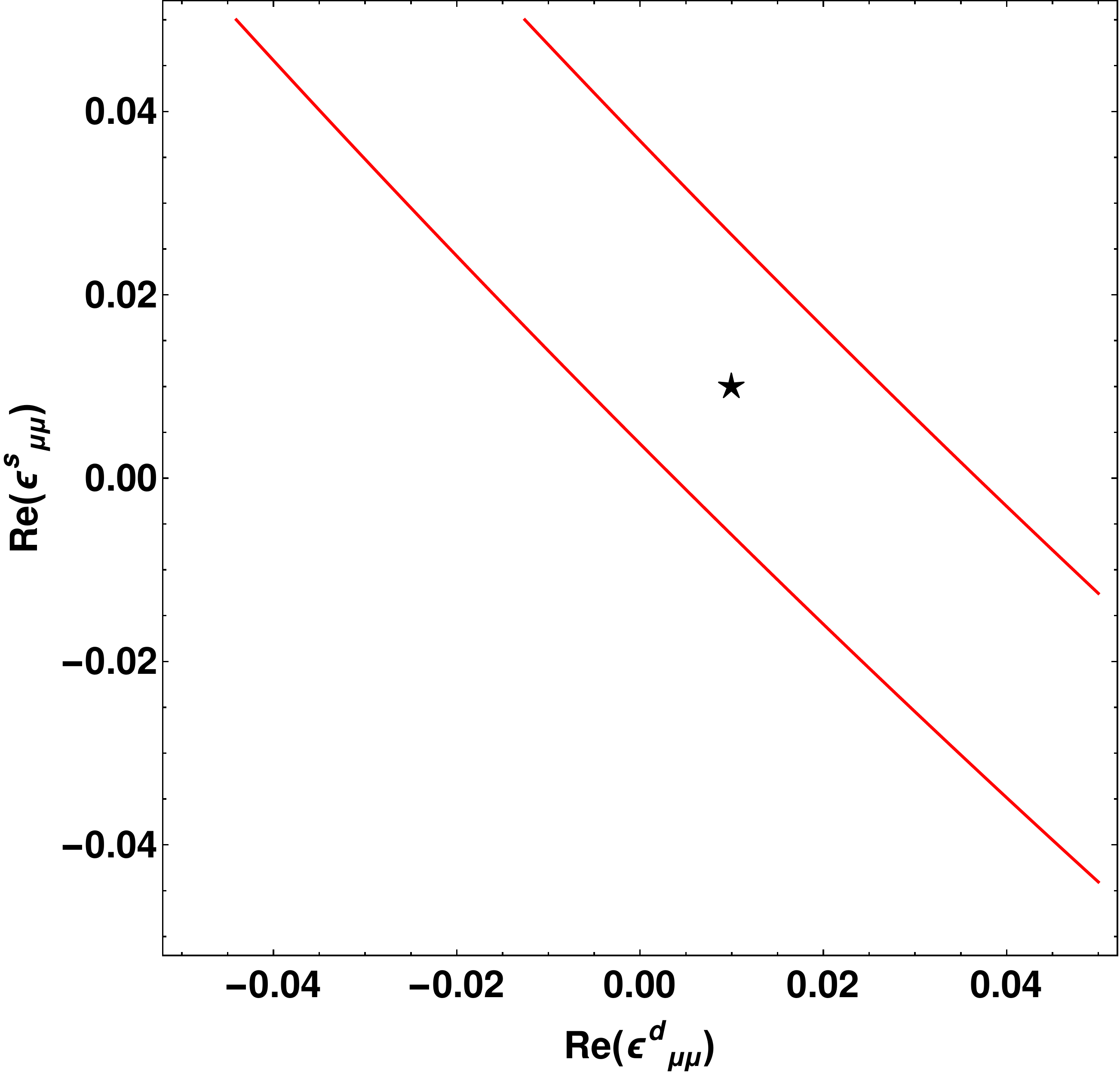}
\includegraphics[height=7cm,width=7cm]{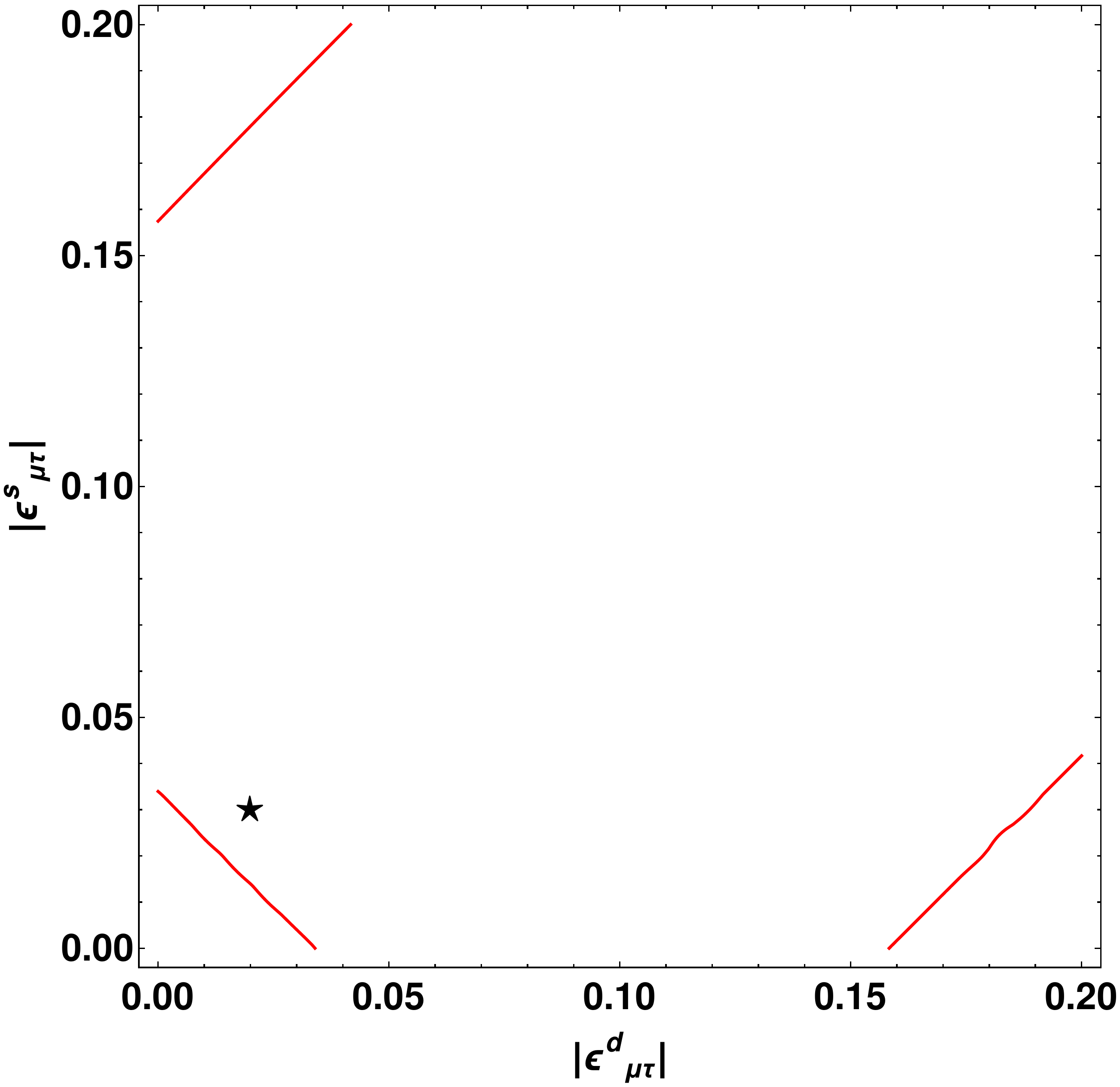}
\caption{\it 90\% CL allowed regions in the measurement of the NSI parameters;  true values are fixed to $\left[\Re(\varepsilon_{\mu\mu}^d),\Re(\varepsilon_{\mu\mu}^s)\right]=(0.01,0.01)$ (left panel) and $\left(|\varepsilon_{\mu\tau}^d|,|\varepsilon_{\mu\tau}^s|\right)=(0.02,0.03)$ (right panel).} 
\label{corr2}
\end{center}
\end{figure}

In the  case of the appearance channel, the $\chi^2$ function reads: 
\begin{equation}
    \chi^2=\frac{N_0^2}{\sigma^2} \left[K_{\alpha\beta}- |\varepsilon_{\alpha\beta}^s|^2-|\varepsilon_{\alpha\beta}^d|^2-2|\varepsilon_{\alpha\beta}^s||\varepsilon_{\alpha\beta}^d|\cos{(\Phi_{\alpha\beta}^s-\Phi_{\alpha\beta}^d})\right] ^2.
\end{equation}
The minima of the $\chi^2$ are always in $\left(\cos{\Delta\Phi}_{min}\right)=\left( \frac{K_{\alpha\beta}-|\varepsilon_{\alpha\beta}^s|^2 -|\varepsilon_{\alpha\beta}^d|^2}{2 |\varepsilon_{\alpha\beta}^s| |\varepsilon_{\alpha\beta}^d|} \right)$;
however, when $|\left(\cos{\Delta\Phi}_{min}\right)|>1$, $\Delta\Phi_{min}$ is forced to be either 0 or $\pi$. Fixing the cut of the $\chi^2$ ($\chi_{0,\alpha\beta}$) at a given CL,
the allowed regions are delimited by:
\begin{eqnarray}
Max\left[0,K_{\alpha\beta}-\sqrt{\frac{ \chi_{0,\alpha\beta}^2 \sigma^2}{N_0^2}}\right]<&\left(|\varepsilon_{\alpha\beta}^s|+|\varepsilon_{\alpha\beta}^d|\right)^2&<K_{\alpha\beta} \nonumber\\ \label{con2} \\ \nonumber  
K_{\alpha\beta}<&\left(|\varepsilon_{\alpha\beta}^s|-|\varepsilon_{\alpha\beta}^d|\right)^2&<K_{\alpha\beta}+\sqrt{\frac{ \chi_{0,\alpha\beta}^2 \sigma^2}{N_0^2}}\,.
\end{eqnarray}
As an example, we report in Fig.(\ref{corr2}) the results of our numerical simulations of the precision achievable in the measurement of the NSI parameters whose true values are fixed to $\left[\Re(\varepsilon_{\mu\mu}^d),\Re(\varepsilon_{\mu\mu}^s)\right]=(0.01,0.01)$ (left panel) and $\left(|\varepsilon_{\mu\tau}^d|,|\varepsilon_{\mu\tau}^s|\right)=(0.02,0.03)$ (right panel).

As we can see, the allowed regions strictly follow the analytic results reported in eqs.(\ref{disallowed}) and (\ref{con2}). In these two examples, data permit to exclude the point (0,0) corresponding to the absence of NSI, but this is not the general case as, for different input values, if $K_{\alpha\beta}<\sqrt{\frac{ \chi_{0,\alpha\beta}^2 \sigma^2}{N_0^2}}$ or $K_{\alpha\alpha}<\sqrt{\frac{ \chi_{0,\alpha\alpha}^2 \sigma^2}{N_0^2}}$, the standard oscillation framework cannot be excluded at the desired confidence level.

\end{document}